\documentclass[preprint,aps,prb]{revtex4}
\usepackage[T1]{fontenc}
\usepackage[latin1]{inputenc}
\usepackage{graphicx}

\begin{document}

\title{Observation of a non-Ohmic Hall resistivity above the critical temperature
in the high-temperature superconductor YBa$_{2}$Cu$_{3}$O$_{7-\delta}$}

\author{I. Puica}

\email{ionut.puica@univie.ac.at}

\author{W. Lang}

\email{wolfgang.lang@univie.ac.at}

\affiliation{Fakultät für Physik, Universität Wien, Boltzmanngasse 5, A-1090 Wien,
Austria}

\author{K. Siraj}

\author{J. D. Pedarnig}

\author{D. Bäuerle}

\affiliation{Institut für Angewandte Physik, Johannes-Kepler-Universität Linz,
A-4040 Linz, Austria}

\begin{abstract}
Investigations of the resistivity and the Hall effect as a function
of electric field and temperature in the normal state and upper part
of the superconducting transition of an optimally doped, very thin
film of YBa$_{2}$Cu$_{3}$O$_{7-\delta}$ are reported. Using a fast
pulsed-current technique allowed to reduce the self-heating of the
sample and to reach electric fields up to 1~kV/cm. An intrinsic non-Ohmic
behavior of the Hall conductivity above the critical temperature that
appears to originate from two different, partially counteracting effects
is revealed. The major contribution stems from the suppression of
Aslamazov-Larkin superconducting fluctuations in high electric fields.
\end{abstract}


\maketitle

\section{Introduction}

The non-linear behavior of the normal-state resistivity close to the
superconducting transition was studied about three decades ago in
the conventional superconductors both theoretically,\cite{Schmid69,Hurault69,Tsuzuki70}
and experimentally on thin aluminum films,\cite{Thomas71,Kajimura71}
and good agreement between experiment and theory was found. The mechanism
for this effect was explained in terms of a change of the paraconductivity
that is evoked by thermodynamic superconducting fluctuations. A sufficiently
high electric field can accelerate the fluctuating paired carriers
so that, on a distance of the order of the coherence length, they
increase their energy by a value corresponding to the fluctuation
Cooper pair binding energy. This results in an additional, electric
field dependent, decay mechanism and leads to deviation of the current-voltage
characteristics from Ohm's law.

In the high-temperature cuprate superconductors (HTSC) many physical
properties including the fluctuation spectrum are significantly different
due to the small coherence lengths and the strongly anisotropic layered
structures of these materials.\cite{Larkin05} Several investigations
of a non-Ohmic in-plane conductivity above $T_{c}$ have been reported
and attributed to the paraconductivity suppression in high electric
fields.\cite{Soret93,Gorlova95,Fruchter04,pssc05} Such measurements
suffer from an inherent self-heating of the samples that needs to
be properly addressed. A satisfactory agreement with the theoretical
models for layered superconductors based on a microscopic approach\cite{Varlamov92}
or on the time-dependent Ginzburg-Landau (TDGL) theory\cite{Mishonov02,PuicaLangE}
has been found only with reduction and correction of the self-heating\cite{Kunchur95}
using a pulsed current technique\cite{Kunchur95,Fruchter04} and
additionally very thin films.\cite{PuicaLangSCST,pssc05}

The influence of high current densities on the Hall effect has been
investigated in HTSC,\cite{Kunchur94,Clinton95,Liebich97,Nakao98,Goeb00}
with the aim of overcoming the vortex pinning and testing its influence
on the Hall anomaly that is observed below $T_{c}$. An intrinsic
non-Ohmic effect above $T_{c}$ was neither explicitly investigated
nor fortuitously found in these studies. A possible reason is that
the applied currents did not exceed $10^{6}$ Acm$^{-2}$, while theoretical
estimates for the suppression of superconducting fluctuations\cite{PuicaLangH}
indicate a rather small effect on the Hall conductivity that might
need even higher current densities to be observed.

In this paper we present our recent investigations of the Hall effect
in YBa$_{2}$Cu$_{3}$O$_{7-\delta}$ (YBCO) above and close to $T_{c}$
in high electric fields, performed with a pulsed-current technique\cite{PuicaLangSCST,pssc05}
that enables significantly higher current densities than have been
previously applied in Hall effect measurements.\cite{Kunchur94,Liebich97}
Our aim is to look for non-Ohmic effects on the Hall conductivity
in the upper part of the superconducting transition region, and to
test whether these effects could be explained as a consequence of
the fluctuation suppression in strong electric fields. The results
can contribute important information to the long-lasting debate on
the origin of the Hall anomaly in HTSC.\cite{Hall04}

\section{Sample preparation and experimental setup}

\label{ExpSetup}In this work we present resistivity and Hall effect
data from an optimally doped $c$-axis-oriented epitaxial YBCO film
with a thickness of 50~nm and with $T_{c}=86.8\,\mathrm{K}$. The
measurements were checked for reproducibility on a second film with
similar results. Our films were prepared by pulsed-laser deposition\cite{bauerle00}
on MgO substrates and patterned by standard photolithography and wet-chemical
etching into a bridge geometry of 200 $\mu$m length and 50 $\mu$m
width with two arms at 100 $\mu$m distance on each side for the voltage
probes. The current contacts were located more than 3~mm apart from
the bridge.

Longitudinal and transverse voltages were collected at the same time
using a pulsed-current technique, whose electric scheme is detailed
elsewhere.\cite{pssc05,PuicaLangSCST} Two identical circuits were
used for the measurement of the longitudinal and transverse voltages.
The electric potentials relative to ground at the sample probes were
connected to the inputs of differential amplifiers with very high
common mode rejection ratio (100000:1 up to $f=100$ kHz, decreasing
proportional to $1/f$ for higher frequencies). Their output signals
were transmitted to the two channels of a high-resolution (14 bit)
digitizer, sampled at a 200 MHz rate, averaged and recorded for further
analysis. A possible deterioration of the transverse voltage signal
due to an insufficient common mode rejection of the differential amplifier
at high frequencies was overcome by the symmetrical arrangement of
the apparatus with respect to ground, with two identical programmable
pulse generators of opposite polarity. The common mode voltage is
thus of the order of the differential signal. The other parts of the
experimental setup consisted of a closed-cycle refrigerator and an
electromagnet. The polarity of the 0.8~T magnetic field was reversed
multiple times for every data point, and for every magnetic field
value the longitudinal and transverse voltage pulses were simultaneously
recorded.

The duration of the square-wave current pulses used in our measurements
was 3.5~$\mu$s, at a repetition frequency of 16 Hz, so that the
cumulative heating between subsequent pulses could be avoided. Moreover,
for pulse duration of the order of $\mu$s, the thermal diffusion
distance in the film is of the order of tens of microns at a quench
velocity of about 10~m/s,\cite{Heinrich05} so that heat generated
at the current contacts (which are a few mm distant from the bridge)
does not interfere with the measurement.

The pulse length is, on the other hand, long enough to establish a
flat plateau of the voltage signal after the initial transient state,
and voltage values could be recorded in a well resolved time-window
of about 1~$\mu$s length in the last third of the pulse duration.
This time-window contained about 200 sampled instant values, whose
average gave the voltage value for the respective pulse. To increase
the signal-to-noise ratio, 1024 subsequent pulses were averaged, allowing
for a measurement precision better than $\pm$1~$\mu$V for the Hall
voltage and better than $\pm$10~$\mu$A for the current. The measurements
were performed at discrete temperature values, measured at the sample
holder with a stability better than $\pm$0.01~K.

The main source of the remaining sample heating during the short high-current
pulses is the limited heat transfer across the film-substrate interface,
which occurs with a characteristic phonon escape time of the order
of 1~ns, $\tau_{es}=R_{b}c_{p}d$, with $d$ the film thickness,
$R_{b}$ the thermal boundary resistance, and $c_{p}$ the phonon
specific heat of the film.\cite{Sergeev94} To overcome this effect
the measurement would have to be performed in sub-ns time after the
current onset, which is out of scope for this kind of measurement.
At the interface to the substrate arises thus an abrupt temperature
difference between an apparent temperature $T_{a}$ measured at the
heat sink (sample holder) and the higher sample temperature $T$:\cite{Kunchur95}
\begin{equation}
T-T_{a}\approx p\, d\, R_{b}\label{DeltaT}\end{equation}
 where $p$ is the dissipated power density in the sample. According
to Eq.~\ref{DeltaT}, the temperature rise can be reduced by a proper
choice of $d$ and $R_{b}$. Hence, we have used as thin as possible
samples that still exhibit a $T_{c}$ close to the bulk value and
MgO as substrate, for its lower thermal resistance as against Al$_{2}$O$_{3}$
or LaAlO$_{3}$.\cite{Sergeev94}

The interface between YBCO and MgO exhibits different kinds of imperfections,
like structural defects and modifications of the local charge density,
which both can reduce the conductivity of the film in the vicinity
of the interface. The intrinsic properties are recovered only at some
distance from the substrate boundary. This defect-rich zone exhibits
a substantially higher resistance, the linear temperature characteristics
of the normal-state resistance is distorted, and $T_{c}$ suppressed.
It typically extends about 10 to 30~nm from the interface into the
YBCO film and has been verified by resistivity measurements and transmission
electron microscopy.\cite{Venkatesan89,Pedarnig02} To account for
this situation in our ultra-thin film we consider an `effective electrical
thickness' of the YBCO film that corresponds to the thickness of the
undistorted top layer. The higher resistivity of our sample as compared
to thicker films prepared under otherwise identical conditions can
be thus ascribed to an effective thickness of the film that is smaller
than the geometrical one. For our 50~nm thick films, an effective
electrical thickness of about 30 nm was assumed. It should be emphasized
that a possible error in the thickness has no influence on the conclusions
regarding a non-Ohmic behavior.

\section{Results and discussion}

\label{Sec-measurements}
\begin{figure}
\includegraphics[width=9cm]{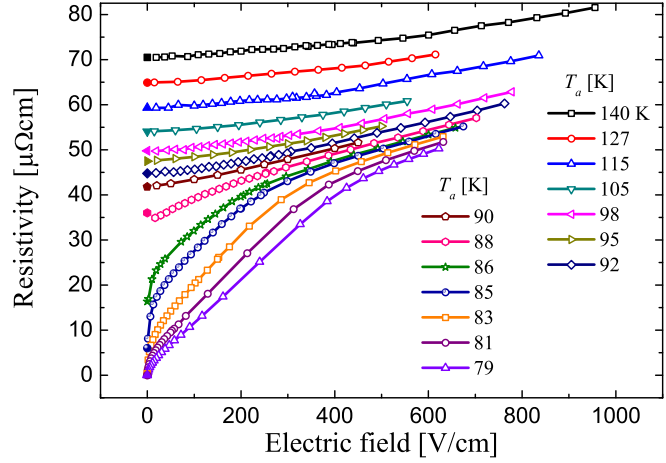}

\caption{\label{RhoXX_EE} (color online). Resistivity of a thin YBCO film
measured with 3.5~$\mu$s current pulses (open symbols and solid
lines) at different fixed sample holder temperatures $T_{a}$ as a
function of the applied electric field. Measurement accuracy is better
than $\pm$0.5~$\mu\Omega$cm (disregarding the uncertainty of the
sample geometrical dimensions, which enter only as scaling factors
in the absolute values of resistivity and electric field). A magnetic
field $B=0.8$~T was applied perpendicular to the film surface. The
filled symbols show the corresponding resistivity values measured
with low DC current.}

\end{figure}

\begin{figure}
\includegraphics[width=9cm]{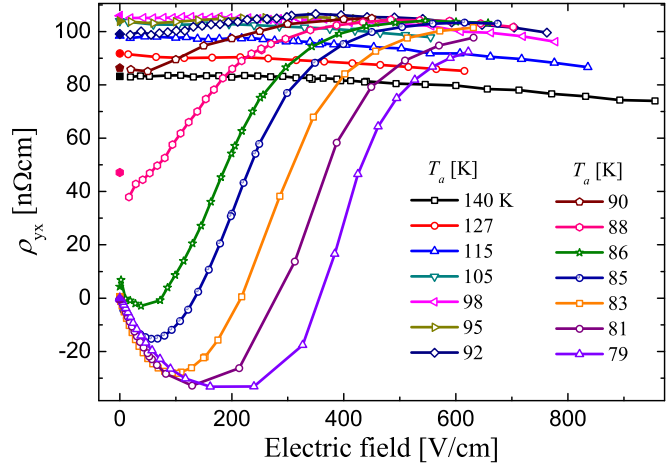}

\caption{\label{RhoYX_EE} (color online). Hall resistivity measured simultaneously
with the resistivity from Fig. \ref{RhoXX_EE} with pulsed currents
(open symbols and solid lines) and with low DC current (filled symbols).
Measurement accuracy is better than $\pm$1~n$\Omega$cm (disregarding
the uncertainty of the sample geometrical dimensions).}

\end{figure}

The longitudinal and the Hall resistivities
of our YBCO thin film are presented in Figs. \ref{RhoXX_EE} and \ref{RhoYX_EE},
respectively. Data were taken with the pulse technique described above
at fixed sample holder temperatures $T_{a}$ with variation of the
injected current and, consequently, the applied electric field. At
very low fields the values measured with the pulse technique approach
the values of a conventional DC measurement (filled symbols), thus,
proving the consistency between the two measurement methods.

The pulse-current technique allowed to attain unprecedented electric
fields of almost 1000 V/cm, current densities as high as 12 MAcm$^{-2}$
(at the rightmost point of the $T_{a}=79\,\mathrm{K}$ curve in Fig.
\ref{RhoXX_EE}) and dissipated power densities of more than 11 GWcm$^{-3}$
(for instance at the highest electric field at $T_{a}=140\,\mathrm{K}$)
without damaging the sample.

\begin{figure}
\includegraphics[width=9cm]{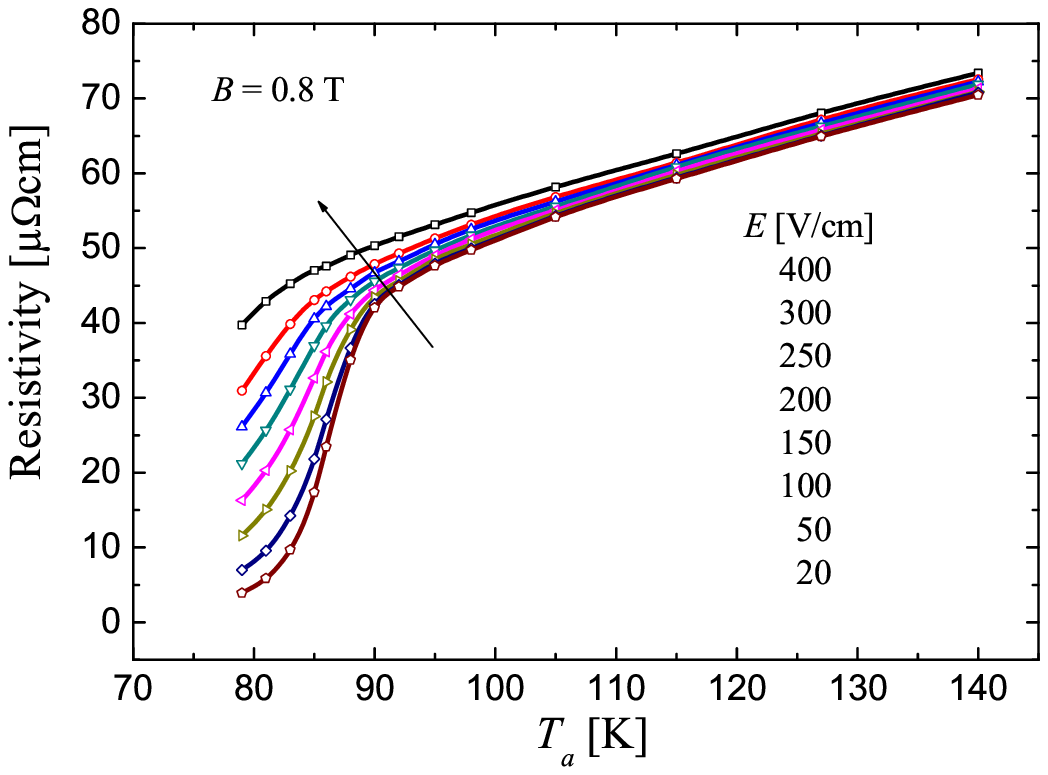}

\caption{\label{RhoXX_Tapp} (color online). Resistivity, derived from data
of Fig. \ref{RhoXX_EE}, as a function of the sample holder temperature
at various constant electric fields. The arrow indicates the sequence
of increasing electric fields.}

\end{figure}

\begin{figure}
\includegraphics[width=9cm]{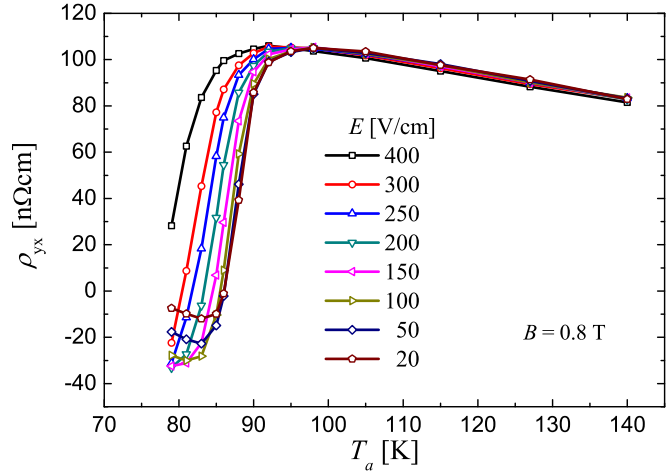}

\caption{\label{RhoYX_Tapp} (color online). Hall resistivity, derived from
data of Fig. \ref{RhoYX_EE}, as a function of the sample holder temperature
at various constant electric fields.}

\end{figure}

However, as one can see from Figs. \ref{RhoXX_EE} and \ref{RhoYX_EE},
a resistivity increase and, respectively, a Hall resistivity decrease
with higher electric fields is present even for temperatures deep
in the normal state, e.g., at $T_{a}=140\,\mathrm{K}$ or $T_{a}=127\,\mathrm{K}$,
where the fluctuation contribution should be vanishingly small. A
non-Ohmic effect from hot electrons like it was observed in many semiconductors\cite{Balkan98}
is unlikely due to the low carrier mobilities in the HTSC. A straightforward
explanation is a temperature increase in the sample, caused by excessive
dissipation and the finite thermal boundary resistance at the film-substrate
interface according to Eq.~\ref{DeltaT}. In order to account for
this self-heating by applying a temperature correction, we first derive
the temperature dependence of the longitudinal and Hall resistivity
at constant electric field from the experimental data, shown in Figs.
\ref{RhoXX_Tapp} and \ref{RhoYX_Tapp}, respectively.

In order to eliminate the spurious effects of self-heating we use
Eq.~(\ref{DeltaT}) to estimate the sample temperature at each individual
data point. The thermal boundary resistance $R_{b}$ is chosen as
a constant fit parameter in such way that all the resistivity curves
superpose in the normal state, up to an uncertainty represented by
the curve thickness. An appropriate value is $R_{b}=0.7\,\mathrm{mKcm^{2}W^{-1}}$,
a value typical for a YBCO/MgO interface.\cite{Nahum91,Sergeev94}
Possible shortcomings of this method could be for instance the assumption
that $R_{b}$ is independent of temperature, temperature drop, and
transferred power density at the film-substrate interface. The thermal
boundary resistance was found indeed to be almost independent on temperature
and heat flux density,\cite{Nahum91,Harrabi00} but also with some
slight decrease with substrate temperature\cite{Sergeev94} and heat
flux\cite{Kelkar97} above $T_{c}$. Since our analysis is limited
to a narrow temperature range such higher-order corrections should
not be important and, if they have an influence at all, they would
increase the non-linearities of the Hall resistivity around 90~K
that are discussed below. Finally, the resistivity and the Hall resistivity
are independent quantities with opposite temperature variations in
the normal state, and the fact that the very same temperature correction
reduces both of them to the low-current characteristics strongly indicates
the self-heating as the only significant effect for nonlinear behavior
at high electric fields deep in the normal state.

\begin{figure}
\includegraphics[width=9cm]{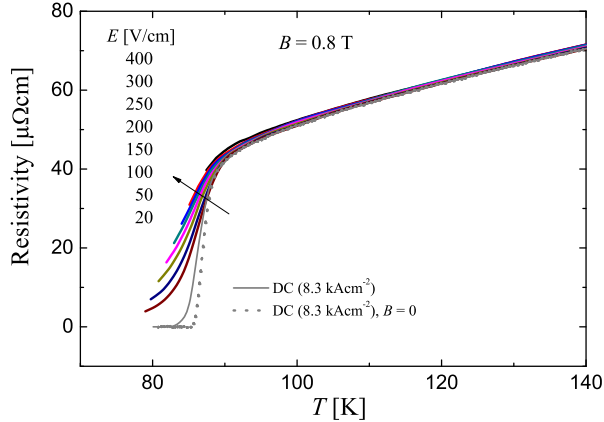}

\caption{\label{RhoXX_Tcorr} (color online). Resistivity data (thick solid lines)
from Fig. \ref{RhoXX_Tapp} according to Eq. \ref{DeltaT}. For reference,
the resistivity measured at a low DC current is also shown, in $B=0.8$~T
(thin solid gray line) and at $B=0$ (dotted gray line). The arrow indicates
the sequence of increasing electric fields. The uncertainty of the
corrected temperature is comparable to the curve thickness.}

\end{figure}

The temperature-corrected resistivity data from Fig. \ref{RhoXX_Tcorr}
exhibit the fan-shape broadening of the superconducting transition
at increasing electric fields, which was also previously experimentally
observed \cite{Kunchur95,Fruchter04,PuicaLangSCST,pssc05} and attributed
to (at least in the high-temperature part of the transition) the suppression
of superconducting fluctuations.\cite{PuicaLangE}

\begin{figure}
\includegraphics[width=9cm]{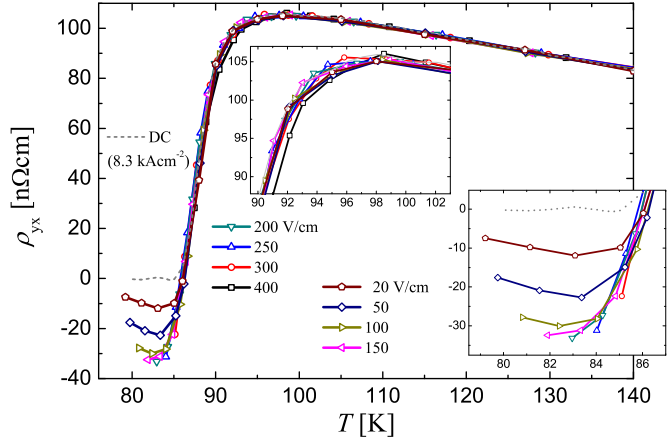}

\caption{\label{RhoYX_Tcorr} (color online). Hall resistivity data from Fig.
\ref{RhoYX_Tapp}, re-scaled by the same temperature correction as
in Fig. \ref{RhoXX_Tcorr}. The dotted curve is the Hall resistivity
measured at low DC current. The insets show details of the Hall resistivity
maximum and its negative minimum, respectively.}

\end{figure}

At first sight, it appears that the Hall resistivity in Fig. \ref{RhoYX_Tcorr}
is Ohmic not only in the normal state, but in the entire region, where
$\rho_{yx}>0$, i.e., down to temperatures where the Hall effect changes
its sign. A weak effect can be seen near the maximum of $\rho_{yx}$,
where the curve changes to a lower slope (see inset of Fig. \ref{RhoYX_Tcorr}).
It is important to realize that the temperature region $T<90$~K,
where a significant non-Ohmic effect appears in the resistivity, corresponds
to the very steep region of the $\rho_{yx}$ curve and, thus, any
small change cannot be seen in the graph. On the other hand, the almost
perfect collapse of the curves in the steep region means that there
is only vanishing uncertainty about the sample temperature, an additional
indication of the validity of our temperature correction procedure.

A very strong non-Ohmic effect can be noticed in the region of $\rho_{yx}<0$,
where the negative minimum is dramatically enhanced at higher electric
fields. A similar behavior has been observed previously --- although
in smaller current densities than in the present work --- and has
been attributed to the fact that pinning forces are overcome in elevated
current densities.\cite{Kunchur94,Liebich97} Our present results
for this temperature region do agree with those conclusions and reveal
additional effects that becomes noticeable in extremely high current
densities. This effects are better evidenced in the Hall conductivity
picture, as it is shown below.

\begin{figure}
\includegraphics[width=9cm]{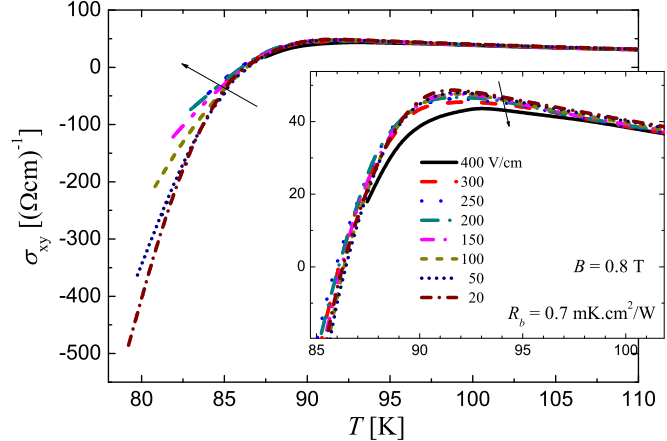}

\caption{\label{SigmaXY_Tcorr} (color online). Hall conductivity computed
from data of Figs. \ref{RhoXX_Tcorr} and \ref{RhoYX_Tcorr}. The
region around the maximum Hall conductivity is shown on a larger scale
in the inset. The arrows indicate the sequence of increasing electric
field. The accuracy of the Hall conductivity is better than $\pm$3~$\Omega^{-1}\mathrm{cm}^{-1}$
(disregarding the uncertainty of the sample geometrical dimensions).
The low-temperature cutoff of the curves is caused by the high-current
limits and the resulting temperature corrections.}

\end{figure}

It has been proposed by several groups that the Hall conductivity
$\sigma_{xy}=\rho_{yx}/(\rho_{yx}^{2}+\rho_{xx}^{2})$ should be appropriate
for an analysis of the Hall effect in the vicinity of $T_{c}$ since
several effects contribute additively and $\sigma_{xy}$ is believed
to be almost independent of pinning.\cite{Vinokur93} This picture
is shown in Fig. \ref{SigmaXY_Tcorr} and, since the curves also have
intrinsically a smaller slope, it allows for a better visualization
of the non-Ohmic Hall effect.

One can now notice two counteracting effects of a high electric field
on the Hall conductivity. The first one is a softening of the drop
of $\sigma_{yx}$ (Fig. \ref{SigmaXY_Tcorr}) that resembles qualitatively
the fan-shape broadening observed for the resistivity (Fig.~\ref{RhoXX_Tcorr}).
Such a non-Ohmic dependence of the Hall conductivity has been recently
theoretically predicted in the frame of the TDGL equation, \cite{PuicaLangH}
based on a suppression of the superconducting fluctuation lifetime.
The behavior displayed in Fig.~\ref{SigmaXY_Tcorr} is in good qualitative
correspondence with the theoretical one, but the magnitude of the
effect is somewhat smaller, possibly also due to the second effect
that will be discussed below.

Some brief remarks on the sign reversal of the Hall effect and its
connection with our present data is appropriate here. The essential
features of the sign reversal and of the Hall anomaly are known to
be governed by a negative contribution to the Hall conductivity that
originates already at $T>T_{c}$ and apparently diverges towards lower
temperatures. It leads to both the sharp drop of the Hall resistivity
and to its sign change.\cite{Rice1,Kang1,Matsuda,Roa01,Clinton95}
The negative contribution to $\sigma_{xy}$ is frequently ascribed,
both by theoretical and experimental work, to fluctuating superconducting
pairs, essentially to those of the Aslamazov-Larkin (AL) process.\cite{Lang94,Samoilov,Smith,Graybeal,Beam}
The Hall resistivity, however, does not diverge towards lower temperatures,
but, on the contrary, vanishes. This can be understood by the fact
that $\rho_{yx}=\sigma_{xy}/\left(\sigma_{xx}^{2}+\sigma_{xy}^{2}\right)\simeq\sigma_{xy}\rho_{xx}^{2}$
is dominated in this temperature range by the behavior of $\rho_{xx}\rightarrow0$.
By the same consideration, the enhancement of the negative Hall anomaly
in Fig.~\ref{RhoYX_Tcorr} turns out to be essentially due to the
non-Ohmic effect on the \emph{longitudinal} resistivity $\rho_{xx}$
that is increased with the electric field, since the change of the
Hall conductivity of Fig.~\ref{SigmaXY_Tcorr}, whose negative value
diminishes in magnitude in high electric fields, would have an opposite
effect on the Hall resistivity. Hence, although a non-Ohmic behavior
of the Hall resistivity can be seen at low temperatures, the effects
under investigation here are obscured in this quantity.

The second effect on the Hall conductivity is better discernible in
the inset of Fig. \ref{SigmaXY_Tcorr}. The maximum of $\sigma_{xy}$
slightly diminishes with increasing fields and this reduction extends
more than $T_{c}+10$~K into the normal state. It is emphasized that
$\sigma_{xy}$ has almost no temperature dependence in this region,
so that an artifact from self-heating or improper temperature correction
can be ruled out. Also, this effect appears to be weaker temperature
dependent than the suppression of AL-type fluctuations discussed before
and has the opposite direction, so that it counteracts the effect
displayed in the main panel of Fig. \ref{SigmaXY_Tcorr}.

We know of no direct prediction of such an effect but would like to
discuss some possible origins. In earlier works on the fluctuation
Hall conductivity under Ohmic conditions an additional Maki-Thompson
(MT) contribution\cite{Maki,Thompson70} was included to obtain a
fit to the experimental data.\cite{Neiman,Lang94} The anomalous
MT fluctuation term cannot be treated in the phenomenological TDGL
theory, and, to our best knowledge, no microscopic theory for the
high electric field effect on the MT contribution is available. It
can be tentatively assumed that the suppression of superconducting
fluctuations by a high electric field would also reduce the MT contribution
to the Hall conductivity. Since the MT term has the same positive
sign as the normal state part and a weaker temperature dependence
as compared to the AL term, such a suppression of MT fluctuations
could evoke the behavior displayed in the inset of Fig. \ref{SigmaXY_Tcorr}.
On the other hand, the \emph{d}-wave pairing symmetry in HTSC is known
to suppress the MT process,\cite{Yip90,Carretta96,Ramallo96,Maki96}
so that the MT contribution to the Hall conductivity and consequently
to its high electric field change could be rather negligible.

Another contribution can result from the reduction of density of states
(DOS) when carriers condense into fluctuating pairs\cite{Larkin05}
or as a result of the pseudogap. The former effect has some similarities
with the MT contribution, but so far has been not explicitly observed
experimentally in the Hall effect of HTSC. A speculative pseudogap
effect can be expected to be rather small in our near-optimally doped
samples, but might be nevertheless worth mentioning, considering that
the observed effect extends only to a few kelvin above $T_{c}$. Clearly,
further theoretical and experimental studies ar needed to provide
possible explanations of this novel effect.

\section{Conclusions}

In summary, we have investigated the non-Ohmic effects on the resistivity
and the Hall effect in optimally doped very thin films of YBCO using
a fast pulsed-current technique. The sample self-heating could be
significantly reduced so that measurements with current densities
of more than $10\,\mathrm{MAcm}^{-2}$ and high electric fields up
to almost 1 kV/cm in the normal state were possible. We have found
evidence of an intrinsic non-Ohmic behavior of the Hall conductivity
above the critical temperature that appears to originate from two
different, partially counteracting effects. One of these could be
identified as the suppression of Aslamazov-Larkin superconducting
fluctuations in high electric fields.

\begin{acknowledgments}
This work was supported by the Austrian Science Fund (FWF) and the
Research Network `Nanoscience and Engineering in Superconductors (NES)'
of the European Science Foundation. We appreciate enlightening discussions
with A. A. Varlamov and T. Mishonov.
\end{acknowledgments}
\bibliographystyle{APSREV}
\bibliography{NonOhmicHall}

\end{document}